\documentclass[aps,twocolumn,showpacs,groupedaddress]{revtex4}  
\pdfoutput=1 
\usepackage{graphicx}  
\usepackage{subfigure} 
\usepackage{dcolumn}   
\usepackage{bm}        
\usepackage{amssymb}   
\usepackage{amsmath}	   
\usepackage{lipsum}		

\begin{document}
\title{Quantum Brownian motion in a magnetic field: Transition from monotonic to oscillatory behaviour}
\author{Urbashi Satpathi}
\address{Raman Research Institute, C. V. Raman Avenue,
Sadashivanagar, Bangalore 560080, India.}
\author{Supurna Sinha}
\address{Raman Research Institute, C. V. Raman Avenue,
Sadashivanagar, Bangalore 560080, India.}
\date{\today}

\begin{abstract}

We investigate the Brownian motion of a charged particle in a magnetic field. 
We study this in the high temperature classical and low temperature quantum domains.
In both domains, we observe a transition of the mean square displacement from a monotonic behaviour to a damped oscillatory behaviour  
as one increases the strength of the magnetic field.
When the strength of the magnetic field is negligible, 
the mean square displacement grows linearly with time 
in the classical domain and logarithmically with time in the quantum domain.
We notice that these features of the mean square displacement are robust and remain essentially the same for an Ohmic dissipation 
model and a single relaxation time model for the memory kernel. 
The predictions stemming from our analysis can be tested against experiments in trapped cold ions. 


\end{abstract}

\pacs{05.40.Jc,05.30.-d,05.40.-a,32.80.Pj,37.10.Ty}
\maketitle

\section{Introduction}
The problem of a Brownian particle suspended in a liquid subject to thermal fluctuations has been studied extensively \cite{doob,Chandrasekhar1,chandra,taylor}.
More recently there has been work in the area of a Brownian particle undergoing diffusion driven by quantum fluctuations
\cite{rafsup,ford,ambegaokar,Leggett}. 

In this paper, we are interested in studying the diffusion behavior of a charged particle in a magnetic field. There have been two approaches towards solving this problem. Leggett et al. \cite{Leggett,leggett2,leggett3} have used the Feynman Vernon path integral approach in which they have solved the dynamics of a charged particle in a magnetic field in the presence of an Ohmic bath. Subsequently, Li et al \cite{ford2, ford2new} have approached the problem via a quantum Langevin equation which corresponds to a reduced description of the system in which the coupling with the heat bath is described by two terms: an operator valued random force $ F(t) $ with mean zero and a mean force characterized by a memory function $ \mu(t) $. In this paper we follow the approach of Ref. \cite{ford2,ford2new} since it is naturally suited to addressing the question of our interest: the time evolution of the mean square displacement of a charged particle in a magnetic field in the presence of viscous dissipation in the high temperature classical domain and the low temperature quantum domain.

While there has been quite a lot of interest in this area \cite{Czopnik,mexicanpaper,ford2,ford2new,cob,shamik,Dattagupta1996,marathe},
most of the studies have focused on the overdamped regime. Here we study in detail the interplay between the 
effect of the magnetic field and damping effects due to dissipation. In particular, a particle of 
charge $q$ and mass $m$ 
in a magnetic field $B$, moves in a circular orbit at a rate set by the cyclotron frequency $\omega_c= qB/{mc}$, where 
$c$ is the speed of light.
The friction coefficient $\gamma$ provides a rate $\gamma^{-1}$ of dissipation. We probe various different
regimes of these two competing time scales both in the high temperature classical domain and the low temperature quantum 
domain and analyze the growth of the mean square displacement in these regimes. 
Furthermore, we suggest an experimental proposal to realize the predictions made in this paper. In particular, to test our predictions experimentally one can proceed as follows \cite{one, two, three}. One can consider cold atom experiments with hybrid traps for ions and neutral atoms and explore the Brownian motion of a charged particle in the presence of a magnetic field induced by Helmholtz coils. 

The paper is organized as follows. In Sec II
we solve the Quantum Langevin Equation for a charged particle in a viscous medium in the presence of a magnetic field \cite{ford2, ford2new}. 
In Sec III we present an analytical expression for the mean square displacement. We then study the high temperature domain and  
probe two regimes - a viscosity dominated regime and a magnetic field dominated regime. 
We do a similar analysis in the low temperature quantum domain. 
We finally end the paper with some concluding remarks in Sec IV.
  
\section{Quantum Langevin equation in the presence of a magnetic field}
The quantum generalized Langevin equation of a charged particle in the presence of a magnetic field is given by \cite{ford2,ford2new}
\begin{eqnarray}
m \ddot{\vec{r}}(t)=-\int_{-\infty}^{t}\mu (t-t')\dot{\vec{r}}(t')dt'+\frac{q}{c}(\dot{\vec{r}}(t)\times\vec{B})+\vec{F}(t)\label{e1}
\end{eqnarray}
where, $ m $ is the mass of the particle, $ \mu(t) $ is the memory kernel, $ q $ is the charge, $ c $ is the speed of light, $ \vec{B} $ is the applied magnetic field and $ \vec{F}(t) $ is the random force with the following properties \cite{ford}
\begin{eqnarray}
\langle F_{\alpha}(t)\rangle &=& 0\label{e2}\\
\frac{1}{2}\langle \left\lbrace F_{\alpha}(t),F_{\beta}(0) \right\rbrace \rangle &=& \frac{\delta_{\alpha\beta}}{2\pi }\int_{-\infty}^{\infty}d \omega \mathrm{Re}\left[ \mu(\omega)\right]\hbar\omega \nonumber\\
&&\mathrm{coth}\left(\frac{\hbar\omega}{2k_{B}T} \right) e^{-i\omega t}\label{e3}\\
\langle\left[ F_{\alpha}(t),F_{\beta}(0)\right]\rangle &=& \frac{\delta_{\alpha\beta}}{\pi}\int_{-\infty}^{\infty}d \omega \mathrm{Re}\left[ \mu(\omega)\right] \nonumber\\
&&\hbar\omega e^{-i\omega t}\label{e4}
\end{eqnarray}
Here, $ \alpha,\beta=x,y,z $, and $ \delta_{\alpha\beta} $ is the Kronecker delta function, such that
\begin{eqnarray*}
\delta_{\alpha\beta} = \begin{cases} 1 &\mbox{if } \alpha = \beta \\ 
0 & \mbox{if } \alpha \neq \beta \end{cases}
\end{eqnarray*}
$ \mu(\omega)=\int_{-\infty}^{\infty}dt \mu(t)e^{i\omega t} $. Eqs. (\ref{e3}) and (\ref{e4}) are obtained from the Fluctuation-Dissipation Theorem which relates the dissipative and fluctuating parts of the quantum Langevin equation (Eq. (\ref{e1})). The dissipative part is characterized by the memory kernel $ \mu(t) $, and the fluctuating part is characterized by the random force $ \vec{F}(t) $.

We assume that the magnetic field is directed along the $ z- $axis, i.e. $ \vec{B}=(0,0,B) $. Then we can write Eq. (\ref{e1}) in terms of 
components as follows:
\begin{eqnarray}
m \ddot{x}=-\int_{-\infty}^{t}\mu (t-t')\dot{x}dt'+\frac{q}{c}\dot{y}B+F_{x}(t)\label{e5}
\end{eqnarray}
\begin{eqnarray}
m \ddot{y}=-\int_{-\infty}^{t}\mu (t-t')\dot{y}dt'-\frac{q}{c}\dot{x}B+F_{y}(t)\label{e6}
\end{eqnarray}
The motion along the $ z- $axis is the same as that of a free particle. The motion along the $ x,y- $axes 
are affected by the magnetic field strength. We restrict our analysis to the $x-y$ plane and study the Brownian motion
of a charged particle in a magnetic field.  

In terms of Fourier transforms, the solutions to these equations are:
\begin{eqnarray}
x(\omega)&=&\frac{1}{m}\frac{i\omega_{c}F_{y}(\omega)-(\omega - iK(\omega))F_{x}(\omega)}{\omega\left[ \omega^{2}-\omega_{c}^{2}-K(\omega)^{2}-2i\omega K(\omega) \right] }\label{e8}\\
y(\omega)&=&\frac{1}{m}\frac{-i\omega_{c}F_{x}(\omega)-(\omega - iK(\omega))F_{y}(\omega)}{\omega\left[ \omega^{2}-\omega_{c}^{2}-K(\omega)^{2}-2i\omega K(\omega) \right] }\label{e9}
\end{eqnarray}

where, $ \omega_{c}=\frac{qB}{mc}$ is the cyclotron frequency, $ K(\omega)=\frac{\mu(\omega)}{m} $. 

\section{Mean square displacement}
The mean square displacement is given by,
\begin{eqnarray}
\langle \Delta r^2 \rangle &=&\langle \left[r(t)-r(0)\right]^2  \rangle \nonumber\\
&=& \langle \Delta x^2 \rangle +\langle \Delta y^2 \rangle 
\end{eqnarray}
\noindent{\bf NOTE: }
The mean square displacement is generally a sum of all the component mean square displacements. In this case, 
we are interested in analyzing the effect of the magnetic field on the growth of the mean square displacement. As the $ z- $component is independent of the magnetic field, we focus only on the $x$ and $y$ components of the mean square displacement.  

The $x$ and $y$ components of the mean square displacement are related to the corresponding position correlation functions as:
\begin{eqnarray}
\langle \Delta x^2 \rangle &=& 2[C_x(0)-C_x(t)]
\end{eqnarray}
Here, $C_x(t)=\frac{1}{2}\langle \left\lbrace x(t),x(0) \right\rbrace \rangle$ is the position correlation function for the $ x $ component. 
Similarly, the $ y $ component of the mean square displacement is related to $ C_y $ as:
\begin{eqnarray}
\langle \Delta y^2 \rangle &=& 2[C_y(0)-C_y(t)]
\end{eqnarray}
$C_y(t)=\frac{1}{2}\langle \left\lbrace y(t),y(0) \right\rbrace \rangle$. 
Using the force-force correlation (Eq. \ref{e3}), we can write the position correlation function for the $ x $ component as follows:
\begin{widetext}
 \begin{eqnarray}
 C_{x}(t)
 =\frac{\hbar}{2\pi m}\int_{-\infty}^{\infty}d\omega \mathrm{Re}[K(\omega)] \frac{\left[\left(\omega + \mathrm{Im}[K(\omega)]\right)^{2}+\omega_{c}^{2}+\mathrm{Re}[K(\omega)]^{2} \right]\mathrm{coth}\left(\frac{\hbar\omega}{2k_{B}T} \right)e^{-i\omega t}}{\omega\left\lbrace  \left[\left(\omega + \mathrm{Im}[K(\omega)]\right)^{2}+\omega_{c}^{2}+\mathrm{Re}[K(\omega)]^{2}  \right]^{2}- 4\omega_c^{2}\left(\omega + \mathrm{Im}[K(\omega)]\right)^{2} \right\rbrace } \label{e10}
 \end{eqnarray}
 \end{widetext}
The same expression is obtained for $ C_{y}(t) $, i.e. $ \langle \Delta x^2 \rangle =\langle \Delta y^2 \rangle $.
The above expressions are obtained by using the fact that the cross correlations of the force components vanish (Eq. (\ref{e3})).
Using the expression for the correlation functions one can write the expression for the 
mean square displacement:
$\langle \Delta r^2 \rangle = \langle \Delta x^2 \rangle +\langle \Delta y^2 \rangle$ as follows:
\begin{widetext}
\begin{eqnarray}
\langle \Delta r^2 \rangle  =\frac{2\hbar}{\pi m}\int_{-\infty}^{\infty}d\omega \mathrm{Re}[K(\omega)] \frac{\left[\left(\omega + \mathrm{Im}[K(\omega)]\right)^{2}+\omega_{c}^{2}+\mathrm{Re}[K(\omega)]^{2} \right]\mathrm{coth}\left(\frac{\hbar\omega}{2k_{B}T} \right)\left(1- e^{-i\omega t}\right) }{\omega\left\lbrace  \left[\left(\omega + \mathrm{Im}[K(\omega)]\right)^{2}+\omega_{c}^{2}+\mathrm{Re}[K(\omega)]^{2}  \right]^{2}- 4\omega_c^{2}\left(\omega + \mathrm{Im}[K(\omega)]\right)^{2} \right\rbrace } \label{e11}
\end{eqnarray}
\end{widetext}
The above expression is an exact expression for the mean square displacement, 
valid for any functional form of the kernel $ K(\omega) $. For examining the interplay 
between the magnetic field and dissipation, we consider a specific kernel. 
The form of the kernel we consider for further analysis is based on the Ohmic 
dissipation model \cite{Leggett, fordtwo}, $ K(t)=2\gamma \delta(t) $. Using this kernel, the mean square displacement
reduces to:
\begin{eqnarray}
\langle \Delta r^2 \rangle&=&\frac{2\gamma\hbar}{\pi m}\int_{-\infty}^{\infty}d\omega  \frac{\left(\omega^{2}+\omega_{c}^{2}+\gamma^{2} \right)}{\omega\left[ \left(\omega^{2}+\omega_{c}^{2}+\gamma^{2} \right)^{2}- 4\omega^{2}\omega_{c}^{2} \right]}\nonumber\\
&&\mathrm{coth}\left(\frac{\hbar\omega}{2k_{B}T} \right)\left(1- e^{-i\omega t}\right)   \label{e12}
\end{eqnarray}
This integral can be solved using Cauchy's residue theorem. The memory kernel satisfies causality which implies,
\begin{eqnarray*}
K(t)=0,& t<0
\end{eqnarray*}
We choose the contour in the lower half plane. The numerator of the integrand has poles in the lower half plane corresponding to $\omega =-in\pi\Omega_{th}$, where $ n $ is a positive integer and $ \Omega_{th}=\frac{2k_B T}{\hbar} $. The denominator has poles at $ \omega= \pm \omega_c \pm i\gamma $. Out of these only two lie in the lower half plane. These poles are at $ \omega= \pm \omega_c - i\gamma $.
The integral in Eq. (\ref{e12}) is $ (-2\pi i) $ times the sum of the residues at the poles. The details of the calculations are given in the Appendix.
\begin{widetext}
\begin{eqnarray}
\langle \Delta r^2 \rangle&=& \frac{\hbar}{\pi m\left( \gamma^{2}+\omega_{c}^{2}\right)}\left\lbrace \left(\gamma+ i\omega_{c} \right) \left[ H_{\frac{\gamma -i \omega _c}{\pi  \Omega_{th} }}+H_{-\frac{\gamma -i \omega _c}{\pi  \Omega_{th} }}\right] 
+\left(\gamma- i\omega_{c} \right)\left[ H_{-\frac{\gamma +i \omega _c}{\pi  \Omega_{th} }}+H_{\frac{\gamma +i \omega _c}{\pi  \Omega_{th} }}\right]\right.\nonumber\\ 
&+&e^{-\pi  t \Omega_{th} } \left(\gamma+ i\omega_{c} \right)\left[ \Phi \left(e^{-\pi  t \Omega_{th} },1,\frac{\gamma +\pi  \Omega_{th} -i \omega
   _c}{\pi  \Omega_{th} }\right)+\Phi \left(e^{-\pi  t \Omega_{th} },1,\frac{-\gamma +\pi    \Omega_{th} +i \omega _c}{\pi  \Omega_{th} }\right)\right]\nonumber\\
&+&e^{-\pi  t \Omega_{th} } \left(\gamma- i\omega_{c} \right)\left[ \Phi \left(e^{-\pi  t \Omega_{th} },1,\frac{\gamma +\pi 
   \Omega_{th} +i \omega _c}{\pi  \Omega_{th} }\right)+\Phi \left(e^{-\pi  t \Omega_{th} },1,-\frac{\gamma -\pi 
   \Omega_{th} +i \omega _c}{\pi  \Omega_{th} }\right)\right] \nonumber\\
&+&  2\gamma\left[ \pi  t \Omega_{th} +2\mathrm{ln} \left(1-e^{-\pi  t \Omega_{th} }\right)\right] +\pi \left(i\gamma + \omega_{c} \right)\left(1-e^{-t \left(\gamma +i \omega _c\right)}\right) \coth \left(\frac{\omega _c-i \gamma }{\Omega_{th} }\right)\nonumber\\
&+&\left.\pi \left(-i\gamma+ \omega_{c} \right)\left(1-e^{-\gamma  t+i t \omega _c}\right) \coth \left(\frac{\omega _c+i \gamma }{\Omega_{th}
   }\right)\right\rbrace  
\end{eqnarray}
\end{widetext}
We now examine the classical and quantum domains considering the high temperature and the low temperature limits respectively.
\subsection{High temperature limit}
\begin{figure}
\includegraphics[scale=0.36]{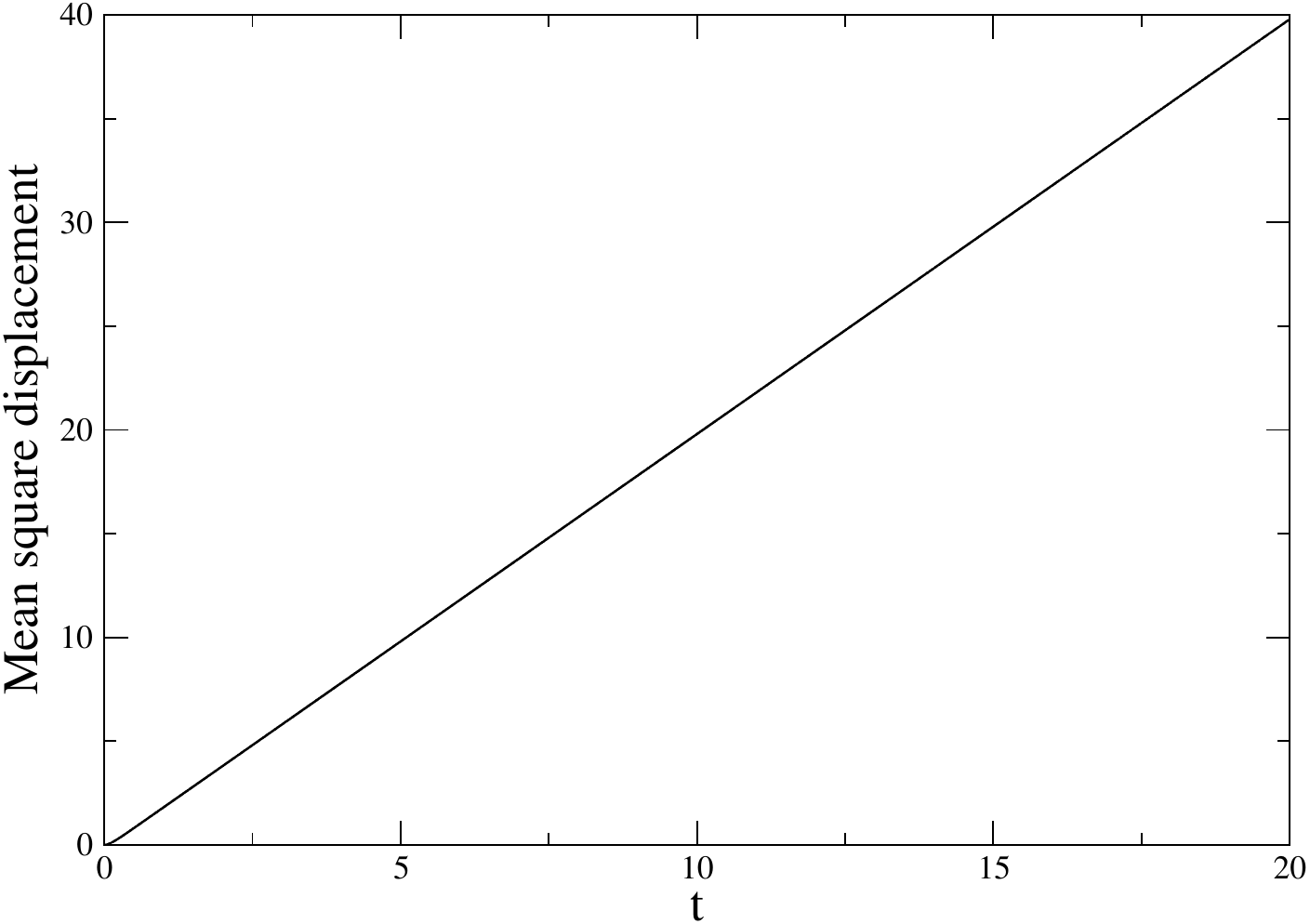}
\caption{\label{fig1}
Plot of the mean square displacement as a function of time $ t $ in arbitrary units under the conditions, $ \frac{\Omega_{th}}{\omega}\gg 1 $ and $\gamma\gg\omega_{c}$. The expression used for this plot is given in Eq. (\ref{eq1}), where we use $ K(t)=2\gamma \delta(t) $. Here we use a scaled time $ t $ where the relaxation time $ \tau_r=\gamma^{-1} $ has been used as the scaling time. 
}
\end{figure}
\begin{figure}
\includegraphics[scale=0.36]{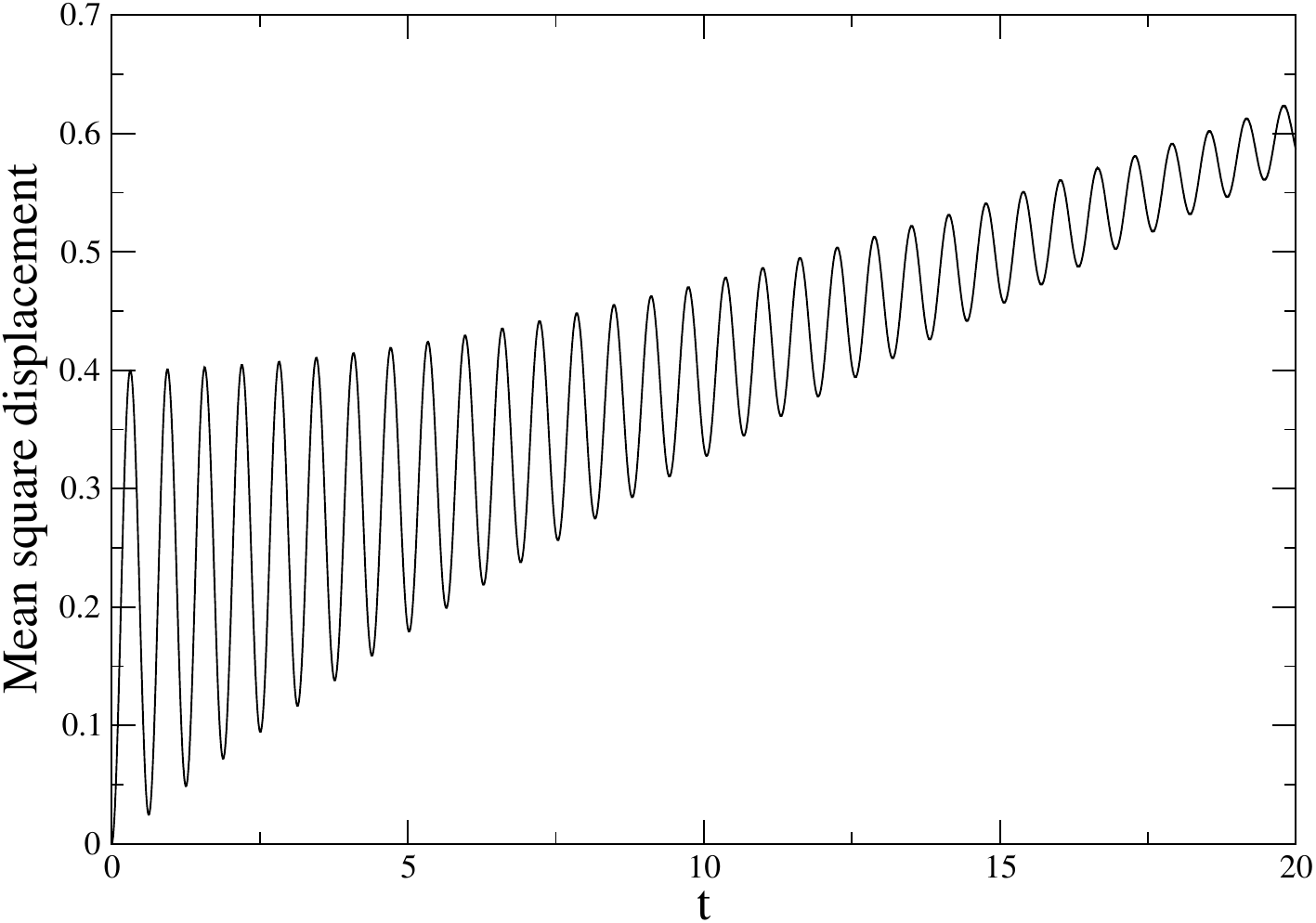}
\caption{\label{fig2}
Plot of the mean square displacement as a function of time $ t $ in arbitrary units under the conditions, 
$ \frac{\Omega_{th}}{\omega}\gg 1 $ and $\omega_{c}\gg\gamma$. The expression used for this plot is given in Eq. (\ref{eq1}), where we use $ K(t)=2\gamma \delta(t) $. Here we use a scaled time $ t $ where the relaxation time $ \tau_r=\gamma^{-1} $ has been used as the scaling time. 
}
\end{figure}
In this limit, $ \frac{\Omega_{th}}{\omega}\gg 1 $. Thus we have:
\begin{eqnarray*}
\mathrm{coth}\left( \frac{\omega}{\Omega_{th}}\right) \sim \frac{\Omega_{th}}{\omega} 
\end{eqnarray*}
The mean square displacement in this limit is:
\begin{eqnarray}
\langle \Delta r^2 \rangle &=&\frac{2\hbar \Omega_{th}}{m}\left\lbrace \frac{\gamma t}{\gamma^{2}+\omega_{c}^{2}} -\frac{\gamma^{2}-\omega_{c}^{2}}{\left( \gamma^{2}+\omega_{c}^{2}\right) ^2}\right.\nonumber\\
&+&\frac{\gamma^{2}-\omega_{c}^{2}}{\left( \gamma^{2}+\omega_{c}^{2}\right) ^2}\mathrm{cos}(\omega_{c}t)e^{-\gamma t}\nonumber\\
&-&\left.\frac{2\gamma\omega_{c}}{\left( \gamma^{2}+\omega_{c}^{2}\right) ^2}\mathrm{sin}(\omega_{c}t)e^{-\gamma t}\right\rbrace \label{eq1}
\end{eqnarray}
This expression is obtained using the asymptotic forms of the Harmonic Number and the Hurwitz-Lerch Transcendent Function. The asymptotic forms and the properties of these functions are discussed in the Appendix.

This case has been discussed in an earlier paper \cite{Czopnik,Dattagupta1996}. 
Here we go beyond Ref. \cite{Czopnik,Dattagupta1996} in investigating in detail 
the competition between the oscillatory effects characterized by the cyclotron 
frequency $ \omega_c $ and the dissipative effects characterized by $ \gamma $. 


Let us consider the case $ \gamma\gg\omega_{c} $. This is the damping dominated regime. In Fig. \ref{fig1} we have shown a plot of the mean square displacement as a function of time $ t $ in this regime. 

Let us consider the case $ \omega_{c}\gg\gamma $. This is the magnetic field dominated regime. In Fig. \ref{fig2} we have shown a plot of the mean square displacement 
as a function of time $ t $ in this regime. Notice that in this domain, the cyclotron effect of the magnetic field on the charged particle leads to 
oscillations in the mean square displacement. This is a qualitatively interesting effect. 

To summarize, we notice a transition from a monotonic to an oscillatory behaviour of the mean square displacement with time as 
the strength of the magnetic field is increased.  
\subsection{Low temperature limit}
\begin{figure}
\includegraphics[scale=0.36]{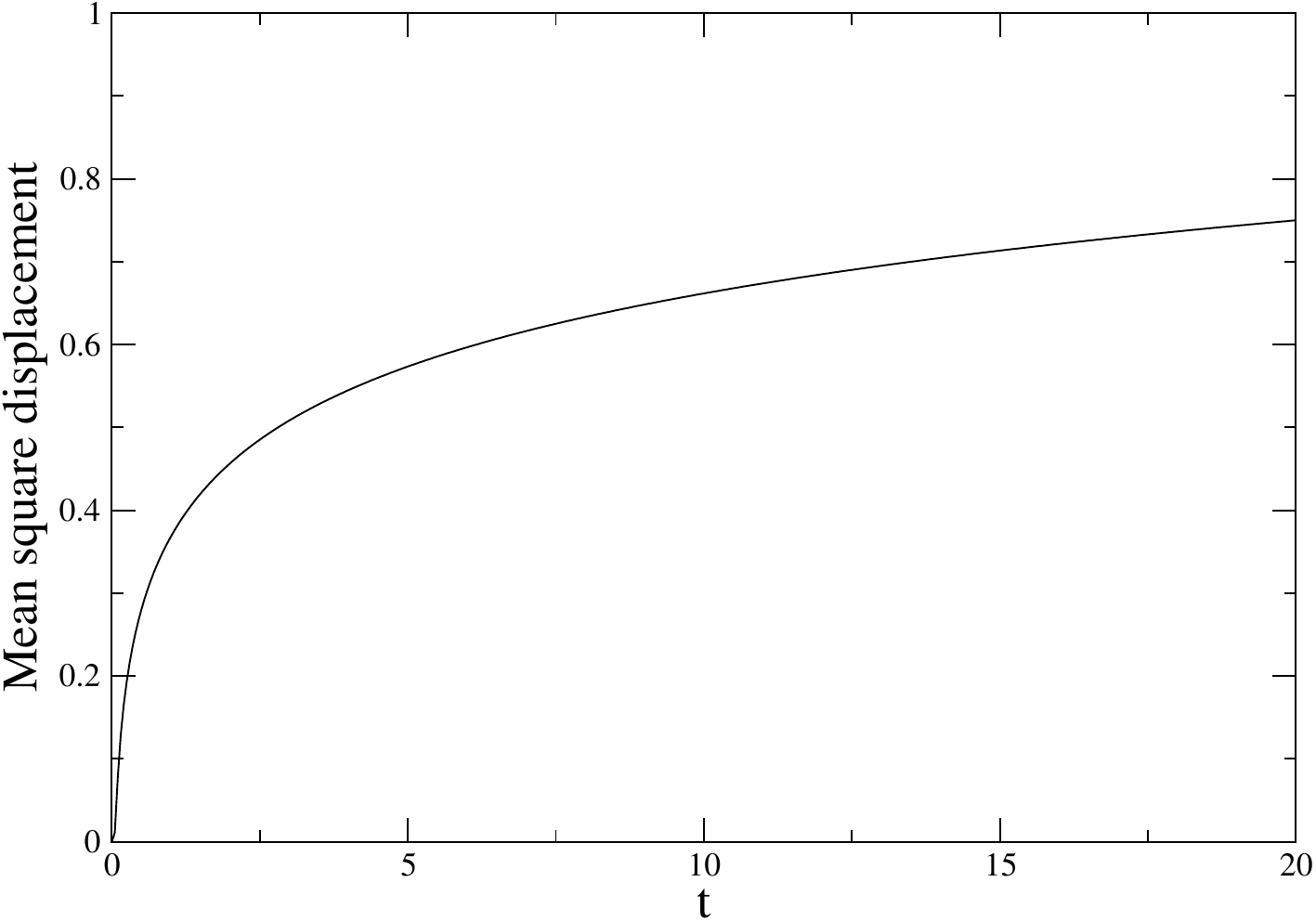}
\caption{\label{fig3}
Plot of the mean square displacement as a function of time $ t $ in arbitrary units under the conditions, $ \frac{\Omega_{th}}{\omega}\ll 1 $ and $\gamma\gg\omega_{c}$. The expression used for this plot is given in Eq. (\ref{eq2}), where we use $ K(t)=2\gamma \delta(t) $. Here we use a scaled time $ t $ where the relaxation time $ \tau_r=\gamma^{-1} $ has been used as the scaling time. 
}
\end{figure}
\begin{figure}
\includegraphics[scale=0.36]{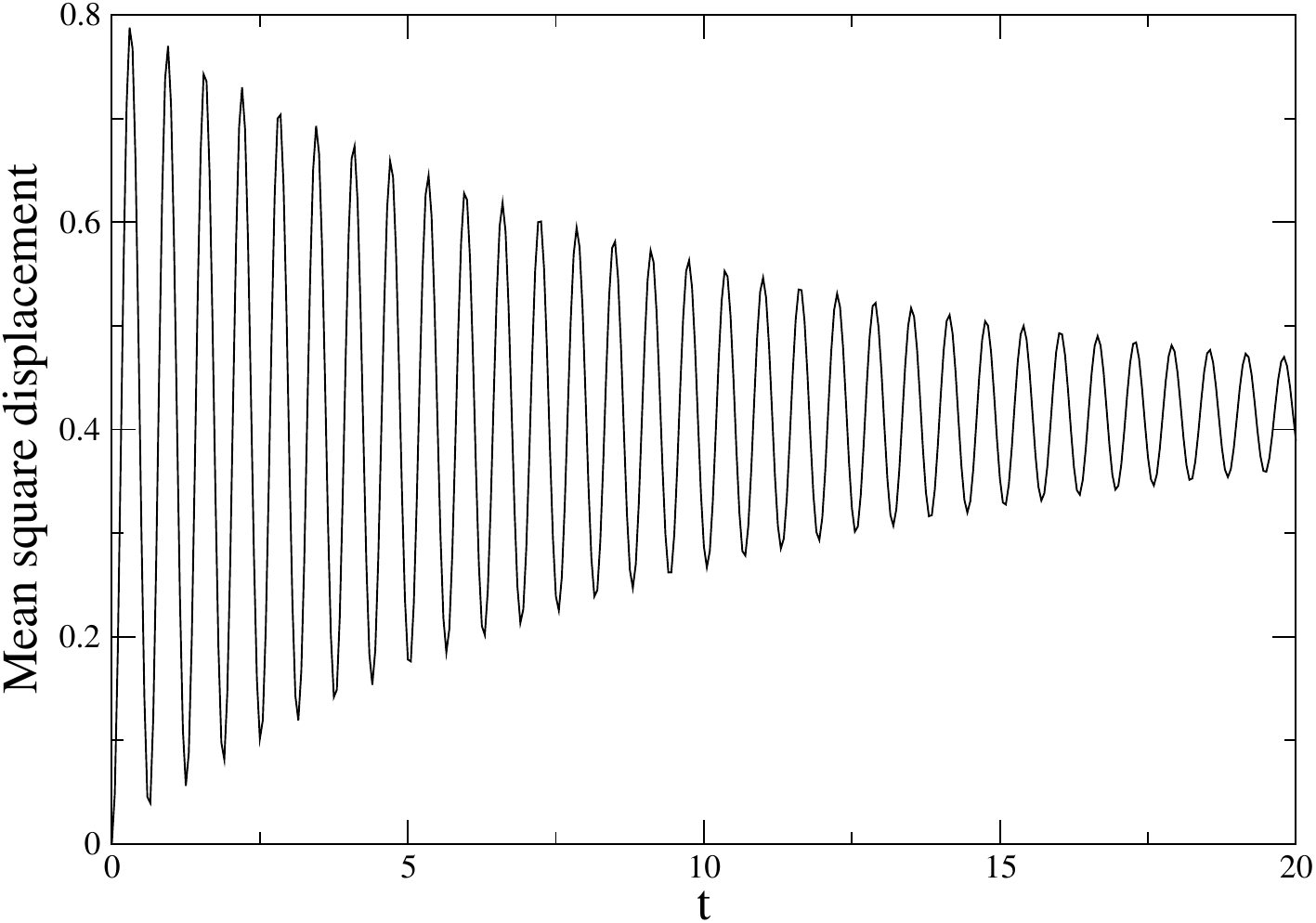}
\caption{\label{fig4}
Plot of the mean square displacement as a function of time $ t $ in arbitrary units under the conditions, $ \frac{\Omega_{th}}{\omega}\ll 1 $ and $\omega_{c}\gg\gamma$. The expression used for this plot is given in Eq. (\ref{eq2}), where we use $ K(t)=2\gamma \delta(t) $. Here we use a scaled time $ t $ where the relaxation time $ \tau_r=\gamma^{-1} $ has been used as the scaling time. 
}
\end{figure}
In this limit, $ \frac{\Omega_{th}}{\omega}\ll 1$. Thus we have:
\begin{eqnarray*}
\mathrm{coth}\left( \frac{\hbar\omega}{2k_{B}T}\right) \sim 1
\end{eqnarray*}
The mean square displacement in this limit is:
\begin{eqnarray}
\langle \Delta r^2 \rangle &=& 
\frac{2\gamma\hbar}{m \pi \left( \gamma^{2}+\omega_{c}^{2}\right) } \left\lbrace 2 \mathrm{ln}\left( (\sqrt{\gamma^{2}+\omega_{c}^{2}})t\right)+2 \gamma_0\right. \nonumber\\
&+&\frac{ \pi\omega_{c}}{\gamma}
-\left. \pi e^{-\gamma t}\left[ \frac{\omega_{c}}{\gamma}\mathrm{cos}(\omega_c t)+ \mathrm{sin}(\omega_c t)\right]  \right\rbrace \label{eq2}
\end{eqnarray}
where, $ \gamma_{0} $ is the Euler-Mascheroni constant.

In contrast to the high temperature limit, the mean square displacement in the low temperature domain is relatively less explored. Here we study the interplay between dissipation and magnetic field in the low temperature domain.

Let us consider the case $ \gamma\gg\omega_{c} $. This is the damping dominated regime for the low temperature domain. In Fig. \ref{fig3} we have shown a plot of the mean square displacement as a function of time $ t $ in this regime. 

Let us consider the case $ \omega_{c}\gg\gamma $. This is the magnetic field dominated regime for the low temperature domain. In Fig. \ref{fig4} we have shown a plot of the mean square displacement as a function of time $ t $ in this regime. 

As in the high temperature regime, in the low temperature regime also we notice a transition from a monotonic to an oscillatory behaviour of 
the mean square displacement with time as the strength of the magnetic field is increased. However, in contrast to the 
high temperature domain, where the monotonic growth of the mean square displacement is linear with time, in the low 
temperature domain the temporal growth of the mean square displacement is logarithmic \cite{rafsup,urbsupraf,marathe}. 

\begin{figure}
\includegraphics[scale=0.36]{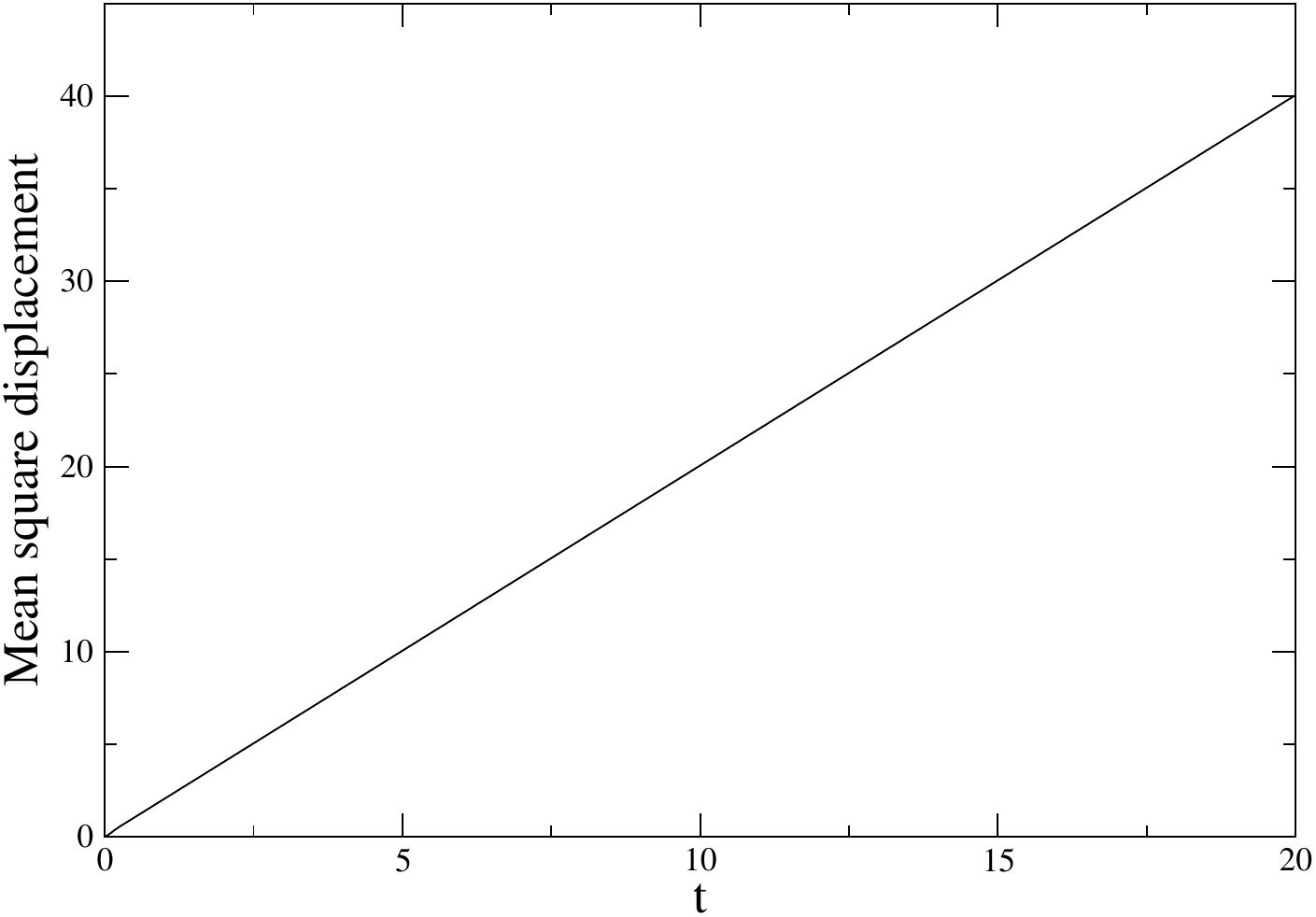}
\caption{\label{fig5}
Plot of the mean square displacement as a function of time $ t $ in arbitrary units under the conditions, $ \frac{\Omega_{th}}{\omega}\gg 1 $ and $\gamma\gg\omega_{c}$, where we use $ K(t)=\frac{\gamma}{\tau} e^{-t/\tau}\theta(t) $. Here we use a scaled time $ t $ where the relaxation time $ \tau_r=\gamma^{-1} $ has been used as the scaling time. 
}
\end{figure}
\begin{figure}
\includegraphics[scale=0.36]{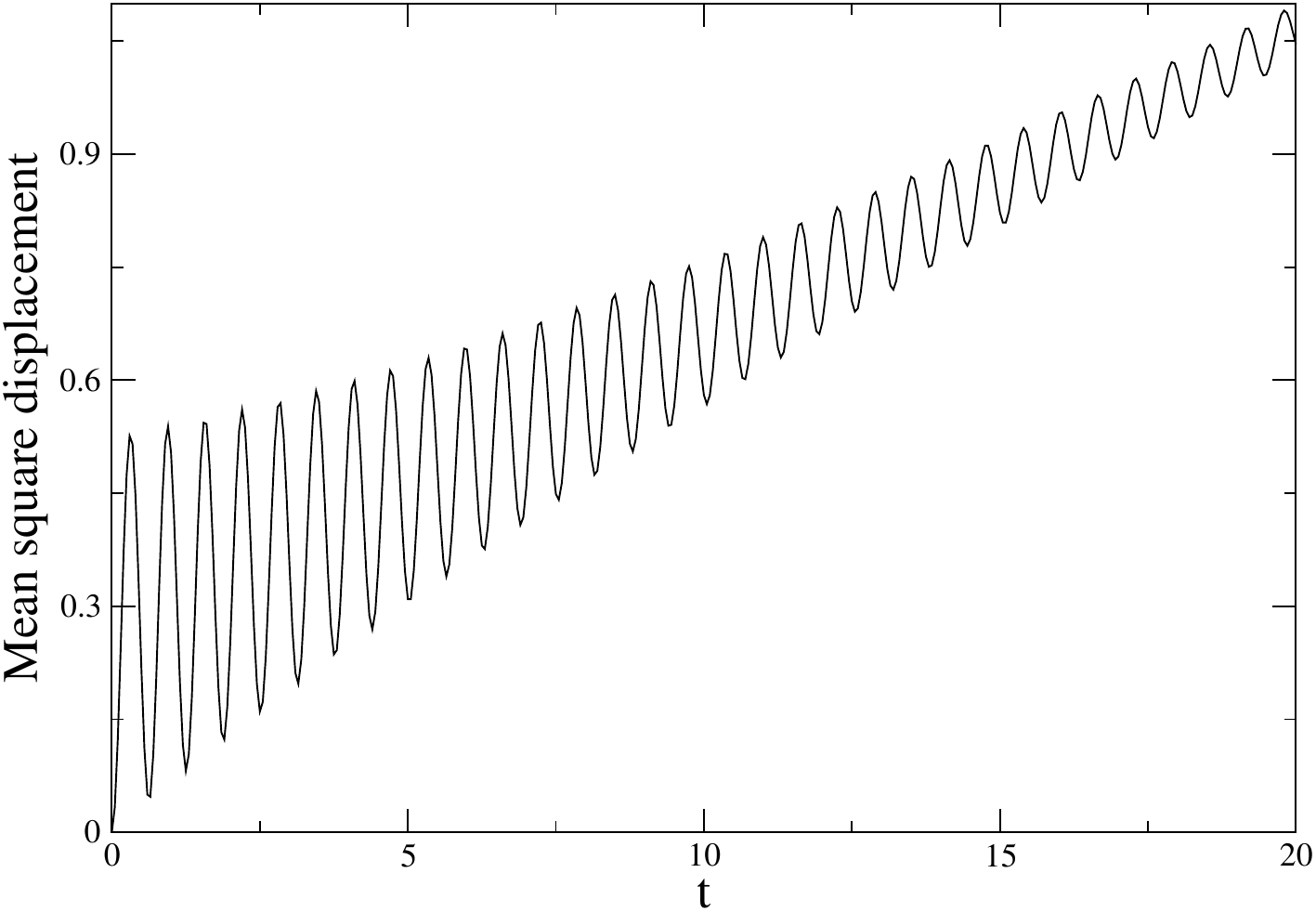}
\caption{\label{fig6}
Plot of the mean square displacement as a function of time $ t $ in arbitrary units under the conditions, $ \frac{\Omega_{th}}{\omega}\gg 1 $ and $\omega_{c}\gg\gamma$, where we use $ K(t)=\frac{\gamma}{\tau} e^{-t/\tau}\theta(t) $. Here we use a scaled time $ t $ where the relaxation time $ \tau_r=\gamma^{-1} $ has been used as the scaling time. 
}
\end{figure}
\begin{figure}
\includegraphics[scale=0.36]{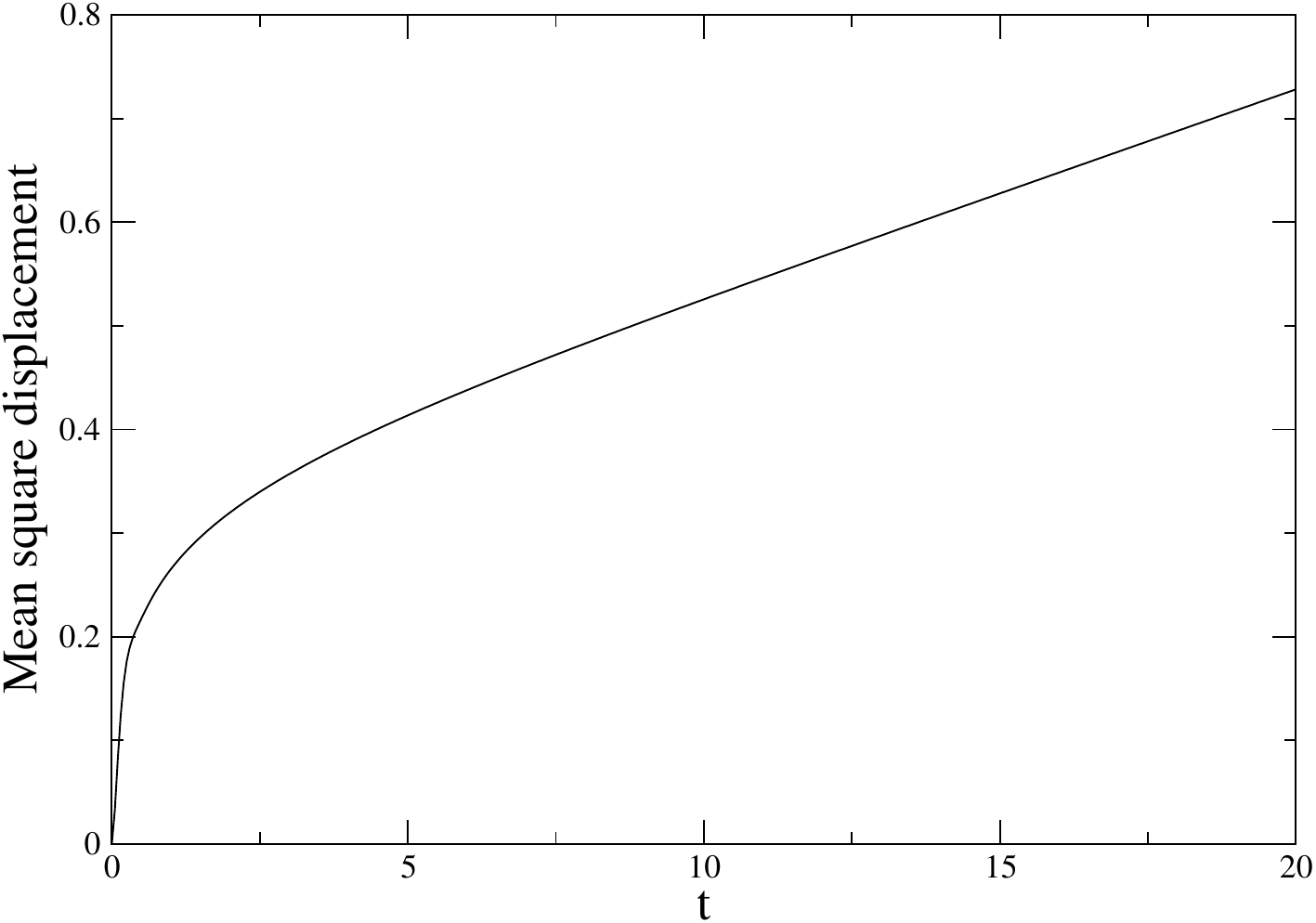}
\caption{\label{fig7}
Plot of the mean square displacement as a function of time $ t $ in arbitrary units under the conditions, $ \frac{\Omega_{th}}{\omega}\ll 1 $ and $\gamma\gg\omega_{c}$, where we use $ K(t)=\frac{\gamma}{\tau} e^{-t/\tau}\theta(t) $. Here we use a scaled time $ t $ where the relaxation time $ \tau_r=\gamma^{-1} $ has been used as the scaling time. 
}
\end{figure}
\begin{figure}
\includegraphics[scale=0.36]{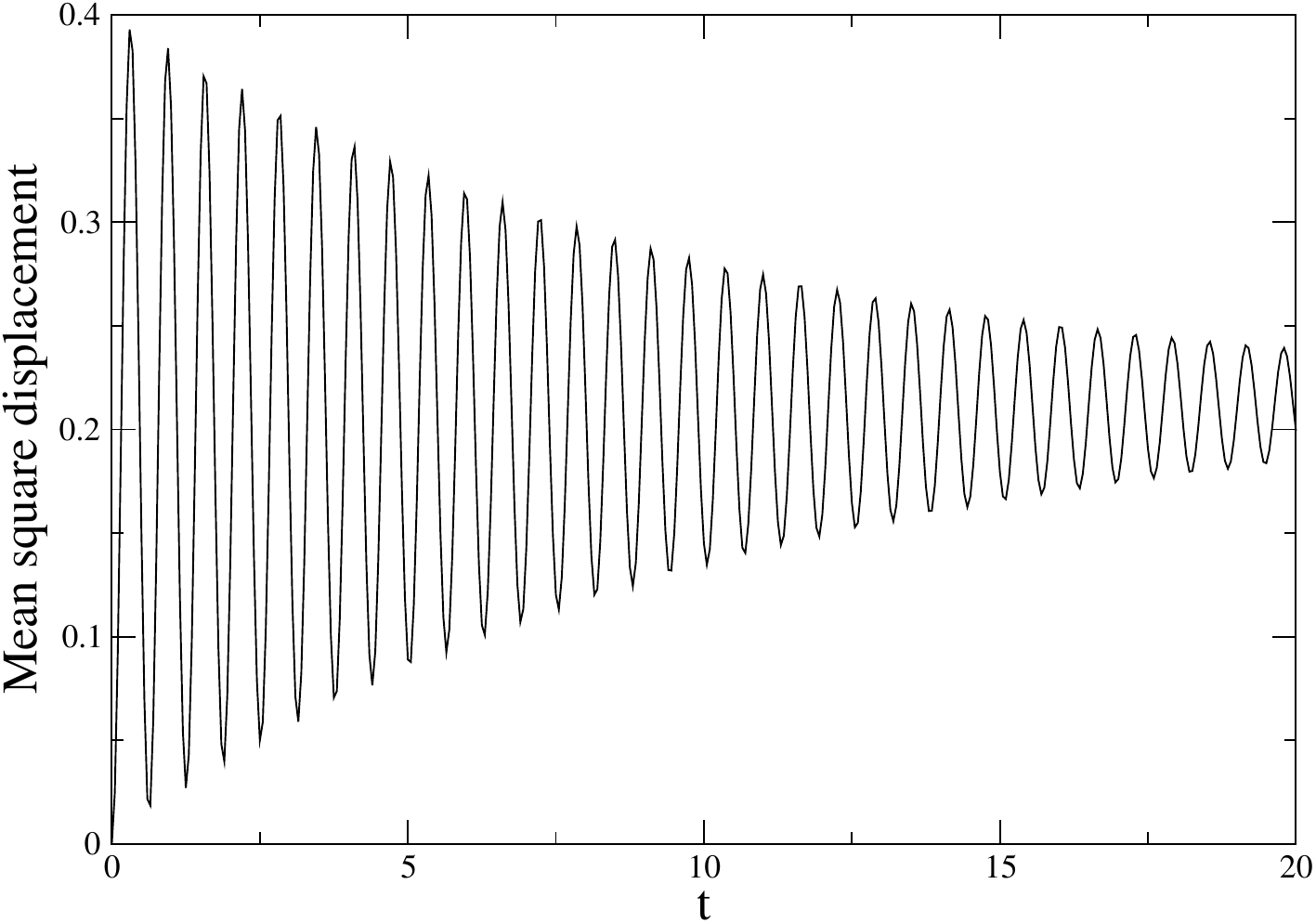}
\caption{\label{fig8}
Plot of the mean square displacement as a function of time $ t $ in arbitrary units under the conditions, $ \frac{\Omega_{th}}{\omega}\ll 1 $ and $\omega_{c}\gg\gamma$, where we use $ K(t)=\frac{\gamma}{\tau} e^{-t/\tau}\theta(t) $. Here we use a scaled time $ t $ where the relaxation time $ \tau_r=\gamma^{-1} $ has been used as the scaling time. 
}
\end{figure}

We have also computed the mean square displacement 
as a function of time for the single relaxation time 
model, where the kernel has the form \cite{fordtwo}
$$K(t) = \frac{\gamma}{\tau} e^{-t/\tau}\theta(t)$$
where $\tau$ is the memory time and $ \theta(t) $ is the Heaviside function. This kernel has been widely used in the literature \cite{marathe, ford1, ford3, maxwell} to model viscous response. 
We notice that the qualitative features of the mean square displacement are essentially the same for the single relaxation time model (see Figs. \ref{fig5}$ - $\ref{fig8}) and the Ohmic model. In other words, the qualitative 
features of the mean square displacement of a charged particle in a viscous medium in the presence of a magnetic field 
are robust and independent of the details of the kernel appearing in the memory function.


\section{Concluding Remarks}
In this paper we have studied the Brownian motion of a charged particle in a magnetic field.
We use the quantum Langevin equation for a charged particle in a magnetic field as a starting point to investigate the 
growth of the mean square displacement as a function of time. 
We analytically derive expressions for the mean square displacement in the high 
temperature classical and low temperature quantum domains using an Ohmic dissipation model for the memory kernel.
In both domains, we notice a transition of the temporal growth of the mean square displacement 
from a monotonic behaviour to a damped oscillatory behaviour  
as one increases the strength of the magnetic field.
We also semi-analytically compute the mean square displacement for a single relaxation time model for the memory kernel and notice
that the qualitative features of the mean square displacement are robust and remain essentially the same for an Ohmic dissipation 
model and a single relaxation time model for the memory kernel. 

The  predictions  for  the  Brownian  motion  of  a  charged  particle  can  be  realized  in  cold  atom 
experiments with hybrid traps for ions and neutral atoms. Such experiments have been realized in 
recent  years  where  a  single  ion  is  immersed  in  a  Bose-Einstein  condensate (BEC) \cite{one}  and  cold 
collisions  between  ions  and  neutral  atoms  are  observed \cite{two}.  In  such  experiments  a  uniform 
magnetic  field  can  be  generated  using  magnetic  coils  in  Helmholtz  configuration.  
The dissipative medium in the high temperature limit can be provided by using a 3D optical molasses \cite{three} combined with a deep optical or magnetic trap.
The temperatures in the low temperature limit can be obtained with BEC in a very shallow trap. 
The inelastic collisions between the single ion 
and the surrounding ultra-cold neutral atoms provide a dissipative medium for the ion. Hence, the 
Brownian  motion  of  a  charged  particle  in  a  magnetic  field  can  be  experimentally  realized  both  in 
the classical and quantum regimes.

\section{Acknowledgements}

We thank Sanjukta Roy for discussions related to the experimental realization of our theoretical predictions.

\renewcommand{\theequation}{A-\arabic{equation}}
  \setcounter{equation}{0}  
\section{Appendix} \label{A}
\begin{widetext}
\begin{eqnarray}
\langle \Delta r^2 \rangle&=&\frac{2\gamma\hbar}{\pi m}\int_{-\infty}^{\infty}d\omega  \frac{\left(\omega^{2}+\omega_{c}^{2}+\gamma^{2} \right)}{\omega\left[ \left(\omega^{2}+\omega_{c}^{2}+\gamma^{2} \right)^{2}- 4\omega^{2}\omega_{c}^{2} \right]}
\mathrm{coth}\left(\frac{\hbar\omega}{2k_{B}T} \right)\left(1- e^{-i\omega t}\right) \label{a1}\\
&=&\frac{i\hbar}{\pi m}\int_{-\infty}^{\infty}d\omega \mathrm{coth}\left(\frac{\hbar\omega}{2k_{B}T} \right)\frac{\left(1- e^{-i\omega t}\right)}{2\omega}
\left\lbrace \frac{1}{\omega +\omega_c + i\gamma}-\frac{1}{\omega +\omega_c - i\gamma} 
+\frac{1}{\omega -\omega_c + i\gamma}-\frac{1}{\omega -\omega_c - i\gamma}\right\rbrace  \nonumber\\
&=& \frac{i\hbar}{\pi m}(I_1-I_2+I_3-I_4) \label{a3}
\end{eqnarray}
Here,
\begin{eqnarray}
I_1=\int_{-\infty}^{\infty}d\omega \frac{\mathrm{coth}\left(\frac{\hbar\omega}{2k_{B}T} \right)\left(1- e^{-i\omega t}\right)}{2\omega\left( \omega +\omega_c + i\gamma\right) }\label{a4}
\end{eqnarray}
The above integral can be evaluated using Cauchy's residue theorem. Since the kernel satisfies causality, the contour where the integrand vanishes is chosen to be a large arc in the lower half plane. The poles are at $ \omega = -in\Omega_{th}\pi, \omega=-(\omega_c + i\gamma) $, where $ n $ is a positive integer and 
\begin{eqnarray}
\Omega_{th}=\frac{2k_B T}{\hbar}
\end{eqnarray}
Therefore, the integral is $(-2\pi i)$ times the residues at the poles, i.e.
\begin{eqnarray}
I_1&=&(-2\pi i)\left\lbrace \frac{H_{-\frac{\gamma -i \omega _c}{\pi  \Omega_{th} }}+e^{-\pi  t \Omega_{th} } \Phi \left(e^{-\pi  t \Omega_{th} },1,\frac{-\gamma +\pi  \Omega_{th} +i   \omega _c}{\pi  \Omega_{th} }\right)+\pi  t \Omega_{th} +\mathrm{ln} \left(1-e^{-\pi  t \Omega_{th} }\right)}{2 \pi  \left(\gamma -i \omega   _c\right)}\right.\nonumber\\
&+&\left.\frac{\left(1-e^{-\gamma  t+i t \omega _c}\right) \coth \left(\frac{\omega _c+i \gamma }{\Omega_{th} }\right)}{2 \left(\omega _c+i \gamma \right)}\right\rbrace 
\end{eqnarray}
Here, $ H $ is the Harmonic Number and $ \Phi $ is the Hurwitz–Lerch Transcendent Function defined respectively as:
\begin{eqnarray*}
 H_x &=&\sum_{k=1}^{x}\frac{1}{k}\\
 \Phi(z,s,\alpha) &=& \sum_{n=0}^{\infty} \frac{z^n}{(n+\alpha)^s}
 \end{eqnarray*} 
The asymptotic forms of the Harmonic Number are:
\begin{eqnarray*}
H_x &=& \begin{cases}
\mathrm{ln}(x)+\gamma_0, &\mbox{if } x\gg 1\\
\frac{\pi^2}{6}x, &\mbox{if } x\ll 1
\end{cases}
\end{eqnarray*}
Here, $ \gamma_0 $ is the Euler-Mascheroni constant. The asymptotic forms of the Hurwitz–Lerch Transcendent Function are discussed in details in ref. \cite{lerch}. 

In Eq. (\ref{a3}), $ I_2 $ is given by,
\begin{eqnarray}
I_2&=&\int_{-\infty}^{\infty}d\omega \frac{\mathrm{coth}\left(\frac{\hbar\omega}{2k_{B}T} \right)\left(1- e^{-i\omega t}\right)}{2\omega\left(\omega +\omega_c - i\gamma\right) }\label{a5}
\end{eqnarray}
In this case, the poles lying in the contour are at $ \omega = -in\Omega_{th}\pi $. Therefore,
\begin{eqnarray}
I_2&=&(-2\pi i)\left\lbrace  -\frac{H_{\frac{\gamma +i \omega _c}{\pi  \Omega_{th} }}+e^{-\pi  t \Omega_{th} } \Phi \left(e^{-\pi  t \Omega_{th} },1,\frac{\gamma +\pi  \Omega_{th} +i   \omega _c}{\pi  \Omega_{th} }\right) +\mathrm{ln} \left(1-e^{-\pi  t \Omega_{th} }\right)}{2 \pi  \left(\gamma +i \omega _c\right)}\right\rbrace 
\end{eqnarray}
In Eq. (\ref{a3}), $ I_3 $ is given by,
\begin{eqnarray}
I_3&=&\int_{-\infty}^{\infty}d\omega \frac{\mathrm{coth}\left(\frac{\hbar\omega}{2k_{B}T} \right)\left(1- e^{-i\omega t}\right)}{2\omega\left(\omega -\omega_c + i\gamma\right) }\label{a6}
\end{eqnarray}
The poles within the contour are at $ \omega = -in\Omega\pi, \omega=\omega_c - i\gamma $. Therefore,
\begin{eqnarray}
I_3 &=&(-2\pi i)\left\lbrace \frac{H_{-\frac{\gamma +i \omega   _c}{\pi  \Omega_{th} }}+e^{-\pi  t \Omega_{th} } \Phi \left(e^{-\pi  t \Omega_{th} },1,-\frac{\gamma -\pi  \Omega_{th} +i \omega _c}{\pi  \Omega_{th}   }\right)+\pi  t \Omega_{th} +\mathrm{ln} \left(1-e^{-\pi  t \Omega_{th} }\right)}{2 \pi  \left(\gamma +i \omega _c\right)}\right.\nonumber\\
&+&\left.\frac{\left(1-e^{-t   \left(\gamma +i \omega _c\right)}\right) \coth \left(\frac{\omega _c-i \gamma }{\Omega_{th} }\right)}{2 \left(\omega _c-i \gamma \right)}\right\rbrace 
\end{eqnarray}
In Eq. (\ref{a3}), $ I_4 $ is given by,
\begin{eqnarray}
I_4&=&\int_{-\infty}^{\infty}d\omega \frac{\mathrm{coth}\left(\frac{\hbar\omega}{2k_{B}T} \right)\left(1- e^{-i\omega t}\right)}{2\omega\left(\omega -\omega_c - i\gamma\right) }\label{a7}
\end{eqnarray}
The poles within the contour are at $ \omega = -in\Omega_{th}\pi $. Therefore,
\begin{eqnarray}
I_4&=&(-2\pi i)\left\lbrace  -\frac{H_{\frac{\gamma -i \omega _c}{\pi  \Omega_{th} }}+e^{-\pi  t \Omega_{th} } \Phi \left(e^{-\pi  t \Omega_{th} },1,\frac{\gamma +\pi  \Omega_{th} -i   \omega _c}{\pi  \Omega_{th} }\right) +\mathrm{ln} \left(1-e^{-\pi  t \Omega_{th} }\right)}{2 \pi  \left(\gamma -i \omega _c\right)}\right\rbrace 
\end{eqnarray}
Collecting expressions for all the integrals:
\begin{eqnarray}
I_{1}-I_{2}+I_{3}-I_{4}&=&\frac{-i}{\left( \gamma^{2}+\omega_{c}^{2}\right) }\left\lbrace \left(\gamma+ i\omega_{c} \right) \left[ H_{\frac{\gamma -i \omega _c}{\pi  \Omega_{th} }}+H_{-\frac{\gamma -i \omega _c}{\pi  \Omega_{th} }}\right] 
+\left(\gamma- i\omega_{c} \right)\left[ H_{-\frac{\gamma +i \omega _c}{\pi  \Omega_{th} }}+H_{\frac{\gamma +i \omega _c}{\pi  \Omega_{th} }}\right]\right.\nonumber\\ 
&+&e^{-\pi  t \Omega_{th} } \left(\gamma+ i\omega_{c} \right)\left[ \Phi \left(e^{-\pi  t \Omega_{th} },1,\frac{\gamma +\pi  \Omega_{th} -i \omega
   _c}{\pi  \Omega_{th} }\right)+\Phi \left(e^{-\pi  t \Omega_{th} },1,\frac{-\gamma +\pi    \Omega_{th} +i \omega _c}{\pi  \Omega_{th} }\right)\right]\nonumber\\
&+&e^{-\pi  t \Omega_{th} } \left(\gamma- i\omega_{c} \right)\left[ \Phi \left(e^{-\pi  t \Omega_{th} },1,\frac{\gamma +\pi 
   \Omega_{th} +i \omega _c}{\pi  \Omega_{th} }\right)+\Phi \left(e^{-\pi  t \Omega_{th} },1,-\frac{\gamma -\pi 
   \Omega_{th} +i \omega _c}{\pi  \Omega_{th} }\right)\right] \nonumber\\
&+&  2\gamma\left[ \pi  t \Omega_{th} +2\mathrm{ln} \left(1-e^{-\pi  t \Omega_{th} }\right)\right] +\pi\left(i\gamma + \omega_{c} \right)\left(1-e^{-t \left(\gamma +i \omega _c\right)}\right) \coth \left(\frac{\omega _c-i \gamma }{\Omega_{th} }\right)\nonumber\\
&+&\left.\pi\left(-i\gamma+ \omega_{c} \right)\left(1-e^{-\gamma  t+i t \omega _c}\right) \coth \left(\frac{\omega _c+i \gamma }{\Omega_{th}
   }\right)\right\rbrace    
\end{eqnarray}
Therefore, the mean square displacement is given by,
\begin{eqnarray}
\langle \Delta r^2 \rangle&=& \frac{\hbar}{\pi m\left( \gamma^{2}+\omega_{c}^{2}\right)}\left\lbrace \left(\gamma+ i\omega_{c} \right) \left[ H_{\frac{\gamma -i \omega _c}{\pi  \Omega_{th} }}+H_{-\frac{\gamma -i \omega _c}{\pi  \Omega_{th} }}\right] 
+\left(\gamma- i\omega_{c} \right)\left[ H_{-\frac{\gamma +i \omega _c}{\pi  \Omega_{th} }}+H_{\frac{\gamma +i \omega _c}{\pi  \Omega_{th} }}\right]\right.\nonumber\\ 
&+&e^{-\pi  t \Omega_{th} } \left(\gamma+ i\omega_{c} \right)\left[ \Phi \left(e^{-\pi  t \Omega_{th} },1,\frac{\gamma +\pi  \Omega_{th} -i \omega
   _c}{\pi  \Omega_{th} }\right)+\Phi \left(e^{-\pi  t \Omega_{th} },1,\frac{-\gamma +\pi    \Omega_{th} +i \omega _c}{\pi  \Omega_{th} }\right)\right]\nonumber\\
&+&e^{-\pi  t \Omega_{th} } \left(\gamma- i\omega_{c} \right)\left[ \Phi \left(e^{-\pi  t \Omega_{th} },1,\frac{\gamma +\pi 
   \Omega_{th} +i \omega _c}{\pi  \Omega_{th} }\right)+\Phi \left(e^{-\pi  t \Omega_{th} },1,-\frac{\gamma -\pi 
   \Omega_{th} +i \omega _c}{\pi  \Omega_{th} }\right)\right] \nonumber\\
&+&  2\gamma\left[ \pi  t \Omega_{th} +2\mathrm{ln} \left(1-e^{-\pi  t \Omega_{th} }\right)\right] +\pi \left(i\gamma + \omega_{c} \right)\left(1-e^{-t \left(\gamma +i \omega _c\right)}\right) \coth \left(\frac{\omega _c-i \gamma }{\Omega_{th} }\right)\nonumber\\
&+&\left.\pi \left(-i\gamma+ \omega_{c} \right)\left(1-e^{-\gamma  t+i t \omega _c}\right) \coth \left(\frac{\omega _c+i \gamma }{\Omega_{th}
   }\right)\right\rbrace    
\end{eqnarray}
\end{widetext}
\bibliography{supurnareferences}
\end{document}